\documentclass[a4paper, oneside, twocolumn, notitlepage, 10pt]{extarticle_ecoc}
\usepackage{ecoc}
\usepackage{svg}

\DeclareUnicodeCharacter{2264}{\ensuremath{\leq}}
\DeclareUnicodeCharacter{00B1}{\ensuremath{\pm}}
\DeclareUnicodeCharacter{00D7}{\ensuremath{\times}}
\defbibnote{myprenote}{}
\usepackage{acronym}
\acrodef{das}[DAS]{Distributed Acoustic Sensing}
\acrodef{sop}[SoP]{State of Polarization}
\acrodef{ml}[ML]{Machine Learning}
\acrodef{ais}[AIS]{Automatic Identification System}
\acrodef{cw}[CW]{continuous-wave}
\acrodef{pbs}[PBS]{Polarising Beam Splitter}
\acrodef{wdm}[WDM]{Wavelength Division Multiplexing}
\acrodef{dsvdd}[DSVDD]{Deep Support Vector Data Description}
\acrodef{sta}[STA]{Short-Term Average}
\acrodef{lta}[LTA]{Long-Term Average}
\begin{document}
\selectlanguage{english} 
\title{Fully Unsupervised Detection of Physical Contacts on Subsea Cables via State-of-Polarization Monitoring}%

\author{
  Agastya Raj\textsuperscript{(1)},
  Alvaro Doval\textsuperscript{(2)},
  Tian Tian\textsuperscript{(1)},
  Steinar Bjørnstad\textsuperscript{(2)},
  Marco Ruffini\textsuperscript{(1)},
}

\maketitle                 
\begin{strip}
  \begin{author_descr}

    \textsuperscript{(1)} School of Computer Science and Statistics, IRIS Research Group, ADAPT Research Centre, Trinity College Dublin, \textcolor{blue}{\uline{Agastya.Raj@tcd.ie}}
 \textsuperscript{(2)} Tampnet AS, Stavanger, Norway.



  \end{author_descr}
\end{strip}

\renewcommand\footnotemark{}
\renewcommand\footnoterule{}
\vspace{-5mm}
\begin{strip}
  \begin{ecoc_abstract}We present a fully unsupervised Fast-Slow DSVDD detector for continuous State-of-Polarization monitoring on a deployed subsea cable. Trained without event labels, it ranks all five confirmed trawler contacts within the top 13 of 122,174 recordings and surfaces additional corroborated cable-contact events. ©2026 The Author(s)
  \end{ecoc_abstract}
\end{strip}

\footnote{This paper is a preprint of a paper accepted in ECOC 2026 and is subject to Institution of Engineering and Technology Copyright. A copy of record will be available at IET Digital Library}

\section{Introduction}

Subsea fibre-optic cables carry over 97\% of intercontinental data traffic, and trawler fishing and anchor dragging are leading causes of cable damage, with dragged-anchor incidents alone accounting for 30–40\% of offshore cable faults~\cite{icpc2025draggedanchors}. \ac{das} and \ac{sop} fibre sensing technologies mitigate these risks by monitoring vibrations and physical disturbances along the cable~\cite{bjornstadFirstImpactMovement2024}. \ac{das} localises approaching trawlers~\cite{waagaard_experience_2022}, but suffers from saturation effects and limited dynamic range for strong signals such as direct impacts~\cite{lin_analysis_2025}. \ac{sop} monitoring provides a complementary approach~\cite{alexoudisMultiModalFiberSensing2026, vaskinnSensingEarthquakesAerial2026}: it can be extracted directly from existing coherent receivers at no marginal hardware cost, does not saturate during strong motions, is compatible with inline amplifiers~\cite{usmaniSmartSensingGrid2025}, and produces unique signatures for distinct physical impacts~\cite{abdelliRiskyEventClassification2024}.

A recent long-term field trial on the Lowestoft–Lista subsea cable established that \ac{sop} responds to real trawler contacts, anchor drags, and environmental forces over multi-month observation periods~\cite{bjornstadFirstObservationsSubsea2026}. However, this was performed manually by visual correlation of \ac{sop} waveforms, and does not scale to network-wide real-time monitoring. Furthermore, \ac{sop} monitoring through live transmission systems is impacted by environmental and equipment noise, within which physical contacts are rare and subtle. This motivates a \ac{ml} approach over classical thresholding. However, the absence of labelled event data from real field trials has largely confined existing \ac{ml}-based \ac{sop} detection to controlled settings: Supervised classifiers trained on labelled examples achieve high accuracy on short sequences with predefined event types~\cite{abdelliVisionTransformersAnomaly2025, sadighiDeepLearningDetection2025, yangLowcomplexitySOPbasedVibration2025, qinExperimentalDemonstrationBending2026}, and semi-supervised methods relax this to one-class fits on labelled baselines~\cite{sadighiMLBasedStatePolarization2026}, but both evaluate on controlled testbed scenarios rather than continuous deployed data. Automated detection on deployed cables has been demonstrated in terrestrial settings~\cite{usmaniSmartSensingGrid2025, pellegriniOverviewStatePolarization2025, abdelliPromptOnceManage2026} but relies on supervised training. 


In this work, we transition from manual and supervised approaches to fully unsupervised event detection on continuous long-duration \ac{sop} data from a deployed subsea cable. We use 92 days of continuous \ac{sop} recordings from the Lowestoft–Lista cable (Tampnet, North Sea)-the same system used in our previous field trial~\cite{bjornstadFirstObservationsSubsea2026}, comprising 122,174 one-minute recordings at 44.1 kHz. We evaluate a deep one-class detection model under unsupervised conditions: no trawler labels are used at any stage of training, model selection, or hyperparameter tuning. The model ranks all ground-truth confirmed trawler contacts within the top 13 of the full 122,174 archive, where lower ranks indicate recordings judged more anomalous by the detector. Beyond the 5 logged events, the framework produces additional recordings not  previously identified during manual reviews of \ac{sop} data. Post-hoc review against the \ac{das} and \ac{ais} records confirmed crossings in which the vessel disabled its \ac{ais} transponder during cable crossing. 

These results demonstrate fully unsupervised detection on continuous deployed cable as a low-cost practical approach, while uncovering significant events missed by manual monitoring even under diverse event characteristics.

\begin{figure}[t]
        \centering
       \includegraphics[width=\linewidth]{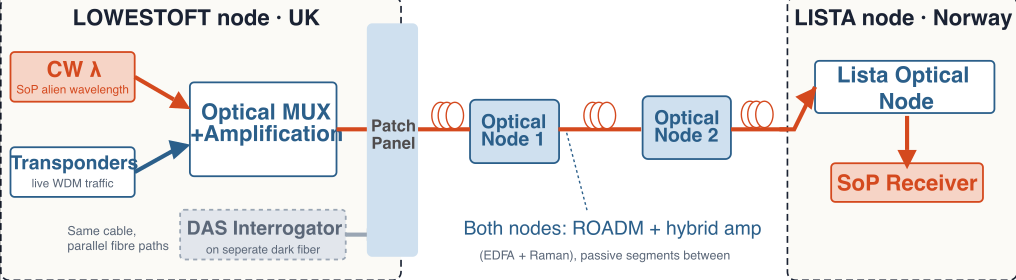}
        \caption{Experimental Setup}
        \label{setup}
        \vspace{-2mm}
\end{figure}
\vspace{-3mm}
\section{Field Trial and Data Pipeline}

\begin{figure*}[t]
    \begin{minipage}[t]{0.5\textwidth}
    \vspace{-12mm}
        \centering
       \includegraphics[width=\linewidth]{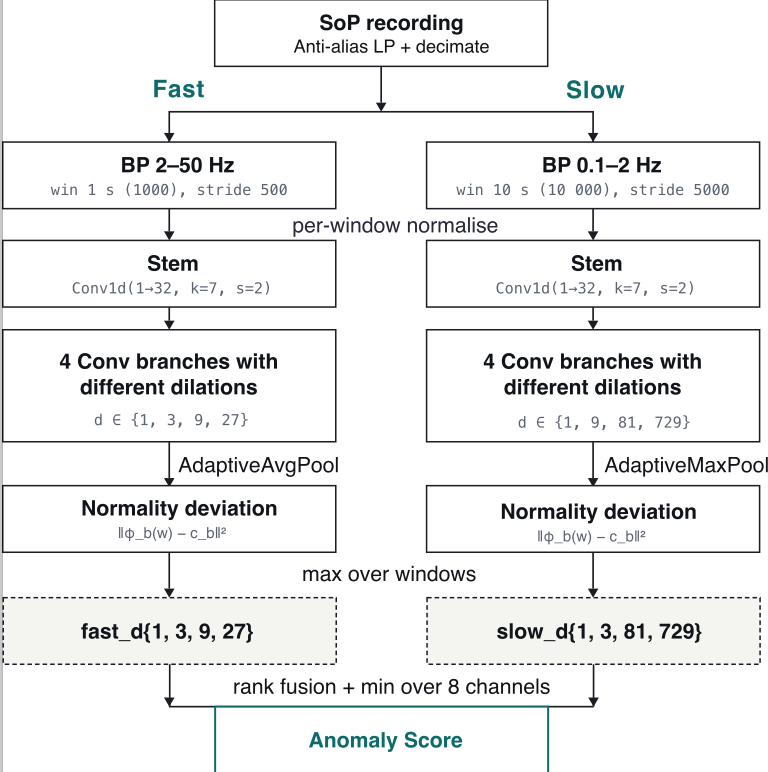}
        \caption{Detection framework of the  Fast-Slow DSVDD model. The input \ac{sop} recording is processed by fast and slow heads, with four dilated convolutional branches. Branch-wise DSVDD distances are fused into a single recording-level anomaly rank.}
        \label{framework}
\end{minipage}\hfill
     \begin{minipage}[t]{0.49\textwidth}
     \vspace{-3mm}
    \centering
    \includegraphics[width=\linewidth]{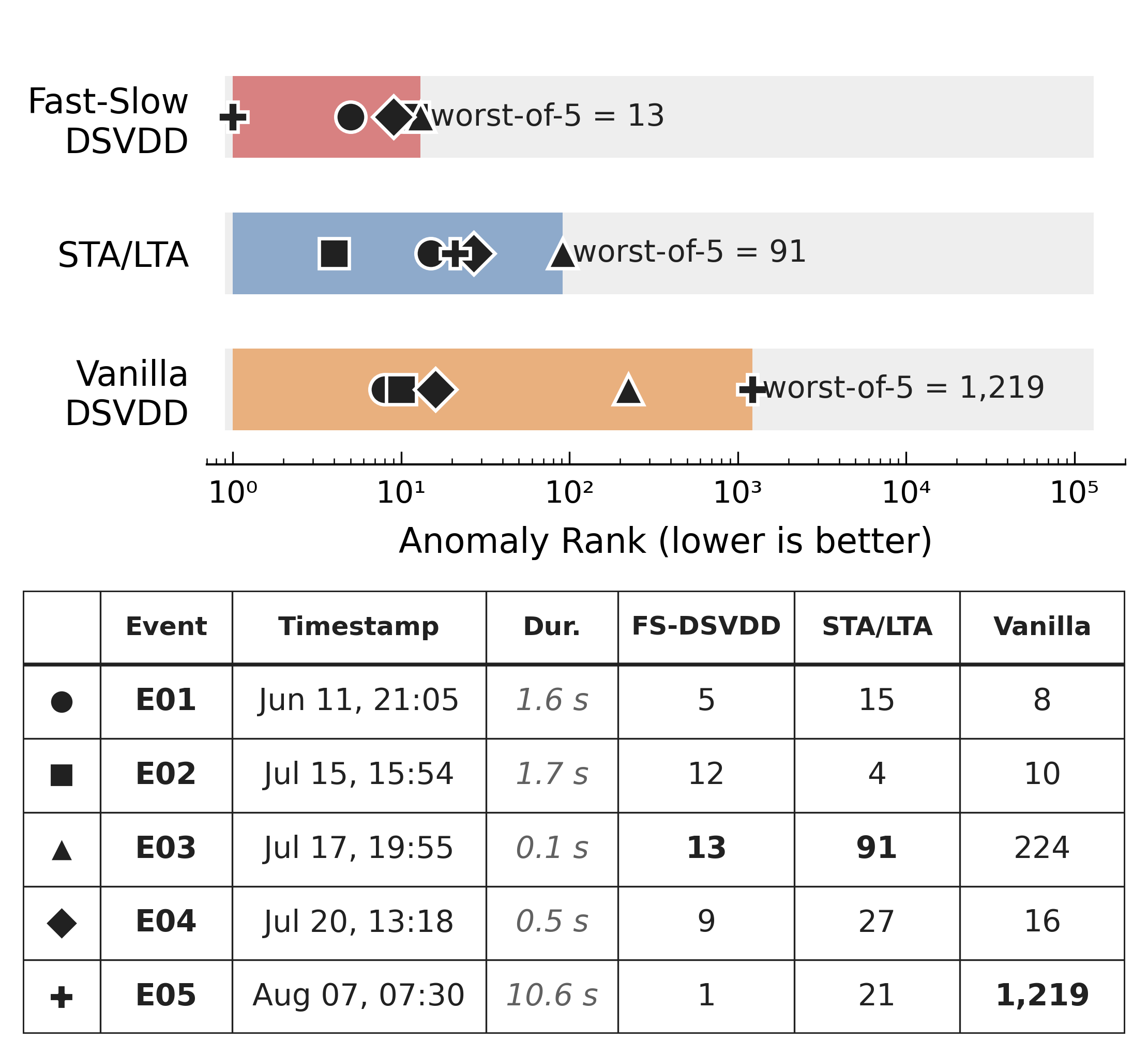}
    \caption{Anomaly ranks of the five confirmed trawler contacts across the 122,174-recording archive. The model scores each recording by how unusual it is relative to normal cable activity; rank 1 corresponds to the most anomalous recording.}
    \label{results}
\end{minipage}
\end{figure*}

Data was acquired from the Tampnet subsea communication system connecting Lowestoft, U.K., to Lista, Norway~(Fig.~\ref{setup}) using the same configuration as ~\cite{bjornstadFirstObservationsSubsea2026}. A \ac{cw} laser was injected as an alien wavelength at Lowestoft, multiplexed with live traffic through an amplified \ac{wdm} transmission system spanning passive submarine cable segments. At Lista, the \ac{cw} wavelength was demultiplexed to a~\ac{pbs} unit measuring the relative $S_1$ Stokes parameter at $44.1$ kHz with 16-bit resolution. \ac{das} was operated on a parallel dark fibre from Lowestoft, covering 120 km of the cable without passing through inline amplifiers; \ac{das} data is not used in the detection framework but provides independent corroboration for events discussed in Section 4.

We analyse 92 days of continuous recordings from this system (June–August 2025), comprising 122,174 one-minute stereo FLAC files. The sensing observable is $S_1 = V_1 - V_2$, the differential output of the two \ac{pbs} channels. During this period, five trawler physical contacts were confirmed by manual cross-referencing of \ac{sop} transients, \ac{das} waterfall signatures, and \ac{ais} vessel-tracking records. These five events, spanning from sub-second (E1, 0.5 s) to sustained (E5, 10.6 s), are summarised in Fig.~\ref{results} and its accompanying table. These five events constitute the sole ground-truth for all detection results; no labels are used to train, select, or tune any model.
\vspace{-2mm}
\section{Detection Framework}

Physical contact events on subsea cables are rare, brief, and unlabelled in operational settings. In our 92-day archive, five confirmed trawler contacts occupy a combined~$\approx$25 seconds of signal across 122,174 one-minute recordings, corresponding to a class prevalence below 0.004\%. Supervised classification is infeasible at this label scarcity, while semi-supervised approaches assume a stationary reference distribution that months of deployed-cable data cannot provide. Standard unsupervised anomaly detectors such as one-class SVMs, Isolation Forests, autoencoders, and \ac{dsvdd}~\cite{dsvdd} operate at a single temporal scale. The \ac{sop} signal in this setting, however, contains two physically different anomaly regimes: trawler contacts produce impulsive transients in the 2--100 Hz band lasting 0.5--10 seconds~\cite{bjornstadFirstObservationsSubsea2026}, while environmental and equipment dynamics produce slower variations below 2 Hz. A single-scale detector must therefore prioritise one regime at the expense of the other. Reliable detection requires a model that learns normal behaviour separately at the characteristic timescales of both regimes.

We extend \ac{dsvdd} to a dual-head architecture, with each head designed for a distinct regime~(Fig.~\ref{framework}). The fast head bandpass filters the input to 2--50 Hz and operates on 1-second windows to capture impulsive transients, while the slow head filters to 0.1--2 Hz and operates on 10-second windows for slower dynamics. Within each head, four parallel dilated convolutional branches provide multi-resolution coverage. The dilation schedules are matched to the regime: the fast head uses ${1,3,9,27}$ and the slow head ${1,9,81,729}$, spanning receptive fields from milliseconds to the full 10-second window. Each branch is associated with a 32-dimensional hypersphere centre, giving eight detection channels. The fast head applies average pooling across 119 overlapping windows, whereas the slow head applies max pooling across 11 windows so that a single anomalous segment can dominate the output. These responses are aggregated into recording-level anomaly scores and fused into a single archive-wide ranking.

Each head is trained on 200,000 uniformly sampled windows from the archive using an unsupervised \ac{dsvdd} objective that encourages embeddings of training windows to lie close to a fixed hypersphere centre. For branch $b$, with encoder $\phi_b$ and centre $c_b$, the anomaly score assigned to a window $x$ is the squared embedding distance $d_b(x) = \| \phi_b(x) - c_b \|^2$, which is minimised during training over unlabelled archive windows using Adam ($lr = 10^{-4}$, 8 epochs). At inference, each branch assigns a recording-level score given by $\max_w d_b(w)$ across that recording's windows. The final anomaly rank of a recording is then defined as the minimum across the eight branch-specific ranks. A recording is therefore flagged if it appears anomalous in any one detection channel.
\vspace{-3mm}
\section{Results}

\begin{figure*}[t]
\vspace{-14mm}
        \centering
       \includegraphics[width=\linewidth]{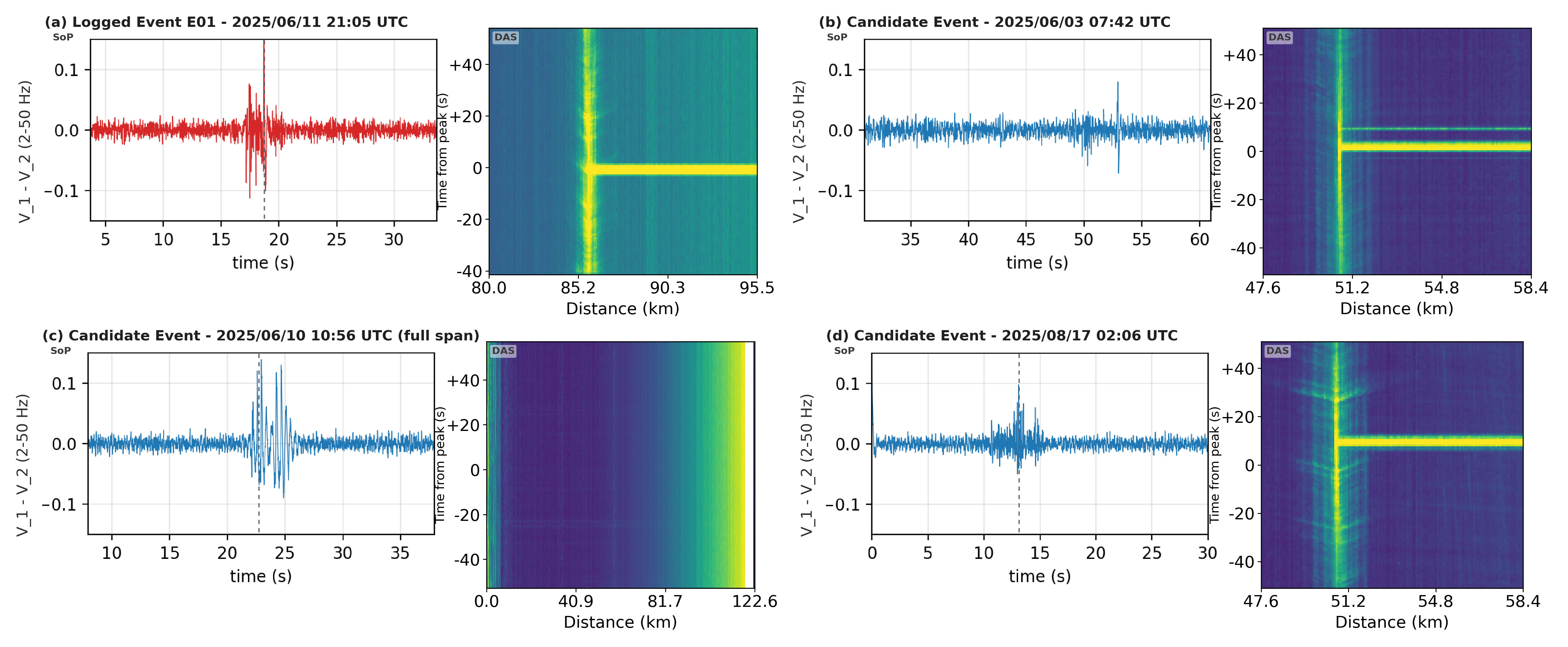}
       \vspace{-6mm}
        \caption{Logged and newly surfaced cable-contact events. Panel (a) shows a logged trawler contact, while panels (b)–(d) show three additional high-ranked candidates from the Fast-Slow \ac{dsvdd} model. In each panel, the left subfigure shows the 30 s \ac{sop} waveform and the right shows the cropped \ac{das} waterfall. Clear \ac{das} signatures appear for the 3 June and 17 August events, while none is observed for the 10 June event within the whole 120 km interrogated section.}
        \label{discovery}
        \vspace{-2mm}
\end{figure*}

We compare Fast-Slow \ac{dsvdd} against two baselines: an \ac{sta}/\ac{lta} trigger (STA = 50 ms, LTA = 5 s, bandpass 0.5--20 Hz on $S_1$), representative of classical transient-detection approaches~\cite{carver_polarization_2024}; and vanilla \ac{dsvdd}, a single-head, single-scale variant of our architecture operating on 1-second windows without the fast-slow band split.

We evaluate each method by sorting the full 122,174-recording archive by anomaly score and reporting the ranks of the five confirmed trawler events. Rank 1 corresponds to the recording the method considers most anomalous, so lower ranks indicate better detection performance. A method that places all five events within the top $K$ requires an operator to review at most $K$ recordings to achieve complete detection. Each ground-truth event is matched to its containing recording using a $\pm 10$ s containment interval around the logged timestamp. We summarise performance using the worst-of-5 metric, defined as the highest anomaly rank among the five confirmed events.

Fig.~\ref{results} and the accompanying table summarise the per-event anomaly ranks together with the corresponding event durations. Fast-Slow \ac{dsvdd} places all five confirmed trawler contacts within the top 13 of the archive, with the sustained 10.6-second event E05 ranked 1st and the sub-second, hardest-to-detect event E03 ranked 13th. \ac{sta}/\ac{lta} achieves worst-of-5 = 91, with its failure concentrated on the shortest events, including E03 at rank 91. Vanilla \ac{dsvdd} achieves worst-of-5 = 1,219, failing on E05 (rank 1,219), the sustained contact that its fixed 1-second window scale cannot resolve. The remaining four events are ranked within the top 30 by both baselines, indicating that simpler methods can recover short impulsive contacts, but not with the same consistency across event durations. What distinguishes Fast-Slow \ac{dsvdd} is consistent performance across the full range of event morphologies, from sub-second impulses to multi-second sustained contacts.

At the operating point where all five trawler contacts are recovered, Fast-Slow \ac{dsvdd} produces 13 alarms over 92 days, approximately one per week. This represents a review burden that is operationally tractable for monitoring. The proposed fast-slow design therefore provides consistent unsupervised detection across both short impulsive and sustained trawler-contact events.

\textbf{Discovery}: The model's highest-ranked recordings included three additional candidate events, on 3\textsuperscript{rd} June, 10\textsuperscript{th} June, and 17\textsuperscript{th} August 2025, that were absent from the event log and had not been identified during manual \ac{sop} review. Figure~\ref{discovery} compares these detections with a logged trawler-contact event. Post-hoc review against concurrent \ac{das} and \ac{ais} records confirmed physical cable-interaction signatures for the 10\textsuperscript{th} June and 17\textsuperscript{th} August events. No corresponding \ac{das} signature was observed for the 10\textsuperscript{th} June event throughout 120 km, suggesting that it occurred outside the \ac{das}-covered cable span in this experiment. These findings show that the proposed framework can surface previously undetected events of physical contact from continuous \ac{sop} monitoring.
\vspace{-2mm}
\section{Conclusions}
We demonstrated fully unsupervised detection of physical contact events on a deployed subsea cable using continuous \ac{sop} monitoring. The proposed Fast-Slow \ac{dsvdd} model recovered all five confirmed trawler contacts within the top 13 of a 122,174-recording archive and also surfaced additional corroborated candidate events. These results establish \ac{sop}-based monitoring as a practical low-cost basis for automated screening of long-duration subsea cable recordings.

\clearpage
\section{Acknowledgements}
This work was supported by the European Union’s Horizon Europe research and innovation programme (Grant No. 10113933) ICON project, CELTIC-NEXT  SUSTAINET-ADVANCED (C2024/3-3), Taighde Éireann – Research Ireland under Grant No. 18/RI/5721 (OpenIreland Research Infrastructure), 13/RC/2106\_P2 (ADAPT centre).

\printbibliography

@online{icpc2025draggedanchors,
  author       = {{International Cable Protection Committee}},
  title        = {Damage to Submarine Cables from Dragged Anchors},
  year         = {2025},
  url          = {https://www.iscpc.org/publications/icpc-viewpoints/damage-to-submarine-cables-from-dragged-anchors/},
  organization = {International Cable Protection Committee},
  note         = {ICPC Viewpoints, updated 24 February 2025, accessed 21 April 2026}
}

@article{bjornstadFirstObservationsSubsea2026,
	title = {First observations of subsea cable physical contacts combining {State} of {Polarization} ({SoP}) and {Distributed} {Acoustic} {Sensing} ({DAS})},
	language = {en},
	author = {Bjørnstad, Steinar and Doval, Alvaro and Tysdal, Anders},
	year = {2026},
}

@inproceedings{bjornstadFirstImpactMovement2024,
	title = {First {Impact} {Movement} {Characterization} of {Shallow} {Buried} {Live} {Subsea}-{Cable}},
	url = {https://ieeexplore.ieee.org/document/10527144},
	urldate = {2026-04-02},
	booktitle = {2024 {Optical} {Fiber} {Communications} {Conference} and {Exhibition} ({OFC})},
	author = {Bjørnstad, Steinar and Yamase Skarvang, Kristina Shizuka and Roar Hjelme, Dag and Tunheim, Asbjørn and Fjermestad, Frode and Østerli, Eivind},
	month = mar,
	year = {2024},
	pages = {1--3},
}

@inproceedings{waagaard_experience_2022,
	address = {Alexandria, Virginia},
	title = {Experience from {Long}-term {Monitoring} of {Subsea} {Cables} using {Distributed} {Acoustic} {Sensing}},
	isbn = {978-1-957171-14-2},
	url = {https://opg.optica.org/abstract.cfm?URI=OFS-2022-Th2.4},
	doi = {10.1364/OFS.2022.Th2.4},
	language = {en},
	urldate = {2026-04-21},
	booktitle = {27th {International} {Conference} on {Optical} {Fiber} {Sensors}},
	publisher = {Optica Publishing Group},
	author = {Waagaard, Ole Henrik and Morten, Jan Petter and Rønnekleiv, Erlend and Bjørnstad, Steinar},
	year = {2022},
	pages = {Th2.4},
}

@article{lin_analysis_2025,
	title = {Analysis of saturation effects of distributed acoustic sensing and detection on signal clipping for strong motions},
	volume = {241},
	copyright = {https://creativecommons.org/licenses/by/4.0/},
	issn = {0956-540X, 1365-246X},
	url = {https://academic.oup.com/gji/article/241/2/971/8063586},
	doi = {10.1093/gji/ggaf089},
	language = {en},
	number = {2},
	urldate = {2026-04-21},
	journal = {Geophysical Journal International},
	author = {Lin, Chen-Ray and von Specht, Sebastian and Ma, Kuo-Fong and Ohrnberger, Matthias and Cotton, Fabrice},
	month = mar,
	year = {2025},
	pages = {971--985},
}

@article{pellegriniOverviewStatePolarization2025,
	title = {Overview on the state of polarization sensing: application scenarios and anomaly detection algorithms},
	volume = {17},
	issn = {1943-0620, 1943-0639},
	shorttitle = {Overview on the state of polarization sensing},
	url = {https://opg.optica.org/abstract.cfm?URI=jocn-17-2-A196},
	doi = {10.1364/JOCN.537881},
	language = {en},
	number = {2},
	urldate = {2026-04-02},
	journal = {Journal of Optical Communications and Networking},
	author = {Pellegrini, Saverio and Minelli, Leonardo and Andrenacci, Lorenzo and Rizzelli, Giuseppe and Pilori, Dario and Bosco, Gabriella and Della Chiesa, Luca and Crognale, Claudio and Piciaccia, Stefano and Gaudino, Roberto},
	month = feb,
	year = {2025},
	pages = {A196},
}

@article{alexoudisMultiModalFiberSensing2026,
	title = {Multi-{Modal} {Fiber} {Sensing} for {Offshore} {Environmental} and {Infrastructure} {Monitoring}},
	language = {en},
	author = {Alexoudis, Konstantinos and Azendorf, Florian and Doval, Alvaro and Bjørnstad, Steinar},
	year = {2026},
}

@article{vaskinnSensingEarthquakesAerial2026,
	title = {Sensing earthquakes in aerial single- and multi-core fiber optic networks},
	volume = {18},
	issn = {1943-0620, 1943-0639},
	url = {https://opg.optica.org/abstract.cfm?URI=jocn-18-4-B119},
	doi = {10.1364/JOCN.584845},
	language = {en},
	number = {4},
	urldate = {2026-04-02},
	journal = {Journal of Optical Communications and Networking},
	author = {Vaskinn, Kristina Skarvang and Elson, Daniel J. and Beppu, Shohei and Soma, Daiki and Bjørnstad, Steinar and Hjelme, Dag Roar and Wakayama, Yuta},
	month = apr,
	year = {2026},
	pages = {B119},
}

@article{usmaniSmartSensingGrid2025,
	title = {A {Smart} {Sensing} {Grid} for {Road} {Traffic} {Detection} {Using} {Terrestrial} {Optical} {Networks} and {Attention}-{Enhanced} {Bi}-{LSTM}},
	volume = {43},
	copyright = {https://ieeexplore.ieee.org/Xplorehelp/downloads/license-information/IEEE.html},
	issn = {0733-8724, 1558-2213},
	url = {https://ieeexplore.ieee.org/document/10891730/},
	doi = {10.1109/JLT.2025.3543180},
	language = {en},
	number = {10},
	urldate = {2026-04-02},
	journal = {Journal of Lightwave Technology},
	author = {Usmani, Fehmida and D'Amico, Andrea and Straullu, Stefano and Aquilino, Francesco and Bratovich, Rudi and Virgillito, Emanuele and Curri, Vittorio},
	month = may,
	year = {2025},
	pages = {4624--4634},
}

@article{abdelliRiskyEventClassification2024,
	title = {Risky event classification leveraging transfer learning for very limited datasets in optical networks},
	volume = {16},
	issn = {1943-0620, 1943-0639},
	url = {https://opg.optica.org/abstract.cfm?URI=jocn-16-7-C51},
	doi = {10.1364/JOCN.517529},
	language = {en},
	number = {7},
	urldate = {2026-04-02},
	journal = {Journal of Optical Communications and Networking},
	author = {Abdelli, Khouloud and Lonardi, Matteo and Gripp, Jurgen and Olsson, Samuel and Boitier, Fabien and Layec, Patricia},
	month = jul,
	year = {2024},
	pages = {C51},
}

@article{yangLowcomplexitySOPbasedVibration2025,
	title = {Low-complexity {SOP}-based vibration broadband sensing and efficient recognition for stable {IM}/{DD} optical interconnects in data centers},
	volume = {17},
	issn = {1943-0620, 1943-0639},
	url = {https://opg.optica.org/abstract.cfm?URI=jocn-17-8-692},
	doi = {10.1364/JOCN.559810},
	language = {en},
	number = {8},
	urldate = {2026-04-02},
	journal = {Journal of Optical Communications and Networking},
	author = {Yang, Bang and Tang, Jianwei and Yu, Huiyang and Hao, Yaguang and Gao, Shuang and Fan, Linsheng and Yao, Yong and Liang, Junpeng and Wei, Jinlong and Yang, Yanfu},
	month = aug,
	year = {2025},
	pages = {692},
}

@article{sadighiMLBasedStatePolarization2026,
	title = {{ML}-{Based} {State} of {Polarization} {Analysis} to {Detect} {Emerging} {Threats} to {Optical} {Fiber} {Security}},
	volume = {23},
	issn = {1932-4537},
	url = {https://ieeexplore.ieee.org/document/11153559/},
	doi = {10.1109/TNSM.2025.3607022},
	urldate = {2026-04-02},
	journal = {IEEE Transactions on Network and Service Management},
	author = {Sadighi, Leyla and Karlsson, Stefan and Natalino, Carlos and Furdek, Marija},
	year = {2026},
	pages = {432--442},
}

@article{qinExperimentalDemonstrationBending2026,
	title = {Experimental {Demonstration} of {Bending} {Eavesdropping} {Detection} in {Optical} {Communications} {Using} a {Physics}-{Informed} {Convolutional} {Network}},
	volume = {44},
	copyright = {https://ieeexplore.ieee.org/Xplorehelp/downloads/license-information/IEEE.html},
	issn = {0733-8724, 1558-2213},
	url = {https://ieeexplore.ieee.org/document/11314660/},
	doi = {10.1109/JLT.2025.3647694},
	language = {en},
	number = {5},
	urldate = {2026-04-02},
	journal = {Journal of Lightwave Technology},
	author = {Qin, Wenshuai and Gong, Xiaoxue and Hou, Weigang and Gan, Lu and Guo, Lei},
	month = mar,
	year = {2026},
	pages = {1636--1646},
}

@article{abdelliPromptOnceManage2026,
	title = {Prompt {Once}, {Manage} {All}: {A} {Unified} {LLM} {Framework} for {Multi}-{Task} {Optical} {Link} {Management}},
	copyright = {https://ieeexplore.ieee.org/Xplorehelp/downloads/license-information/IEEE.html},
	issn = {0733-8724, 1558-2213},
	shorttitle = {Prompt {Once}, {Manage} {All}},
	url = {https://ieeexplore.ieee.org/document/11370165/},
	doi = {10.1109/JLT.2026.3660177},
	language = {en},
	urldate = {2026-04-02},
	journal = {Journal of Lightwave Technology},
	author = {Abdelli, Khouloud},
	year = {2026},
	pages = {1--12},
}

@article{sadighiDeepLearningDetection2025,
	title = {Deep {Learning} for {Detection} of {Harmful} {Events} in {Real}-{World}, {Noisy} {Optical} {Fiber} {Deployments}},
	volume = {43},
	copyright = {https://creativecommons.org/licenses/by/4.0/legalcode},
	issn = {0733-8724, 1558-2213},
	url = {https://ieeexplore.ieee.org/document/10948276/},
	doi = {10.1109/JLT.2025.3557748},
	language = {en},
	number = {13},
	urldate = {2026-04-02},
	journal = {Journal of Lightwave Technology},
	author = {Sadighi, Leyla and Karlsson, Stefan and Natalino, Carlos and Wosinska, Lena and Ruffini, Marco and Furdek, Marija},
	month = jul,
	year = {2025},
	pages = {6092--6101},
}

@article{abdelliVisionTransformersAnomaly2025,
	title = {Vision {Transformers} for {Anomaly} {Classification} and {Localization} in {Optical} {Networks} {Using} {SOP} {Spectrograms}},
	volume = {43},
	copyright = {https://ieeexplore.ieee.org/Xplorehelp/downloads/license-information/IEEE.html},
	issn = {0733-8724, 1558-2213},
	url = {https://ieeexplore.ieee.org/document/10806557/},
	doi = {10.1109/JLT.2024.3519755},
	language = {en},
	number = {4},
	urldate = {2026-04-02},
	journal = {Journal of Lightwave Technology},
	author = {Abdelli, Khouloud and Lonardi, Matteo and Boitier, Fabien and Correa, Diego and Gripp, Jurgen and Olsson, Samuel and Layec, Patricia},
	month = feb,
	year = {2025},
	pages = {1902--1914},
}

@InProceedings{dsvdd,
  title = 	 {Deep One-Class Classification},
  author =       {Ruff, Lukas and Vandermeulen, Robert and Goernitz, Nico and Deecke, Lucas and Siddiqui, Shoaib Ahmed and Binder, Alexander and M{\"u}ller, Emmanuel and Kloft, Marius},
  booktitle = 	 {Proceedings of the 35th International Conference on Machine Learning},
  pages = 	 {4393--4402},
  year = 	 {2018},
  editor = 	 {Dy, Jennifer and Krause, Andreas},
  volume = 	 {80},
  series = 	 {Proceedings of Machine Learning Research},
  month = 	 {10--15 Jul},
  publisher =    {PMLR},
  url = 	 {https://proceedings.mlr.press/v80/ruff18a.html}
}

@article{carver_polarization_2024,
	title = {Polarization sensing of network health and seismic activity over a live terrestrial fiber-optic cable},
	volume = {3},
	copyright = {2024 The Author(s)},
	issn = {2731-3395},
	url = {https://www.nature.com/articles/s44172-024-00237-w},
	doi = {10.1038/s44172-024-00237-w},
	language = {en},
	number = {1},
	urldate = {2026-04-22},
	journal = {Communications Engineering},
	publisher = {Nature Publishing Group},
	author = {Carver, Charles J. and Zhou, Xia},
	month = jul,
	year = {2024},
	pages = {91},
}

\vspace{-4mm}

\end{document}